\documentclass[aps,twocolumn,showpacs,showkeys]{revtex4}

\begin{document}


\newcommand\beq{\begin{equation}}
\newcommand\eeq{\end{equation}}
\newcommand\beqa{\begin{eqnarray}}
\newcommand\eeqa{\end{eqnarray}}
\newcommand\ket[1]{|#1\rangle}
\newcommand\bra[1]{\langle#1|}
\newcommand\scalar[2]{\langle#1|#2\rangle}

\newcommand\jo[3]{\textbf{#1}, #3 (#2)}


\title{\Large\textbf{Unconditionally secure quantum bit commitment using modified double-slit and unstable particles}}

{\large

\author{Chi-Yee Cheung}

\email{cheung@phys.sinica.edu.tw}

\affiliation{Institute of Physics, Academia Sinica\\
             Taipei, Taiwan 11529, Republic of China\\}


\begin{abstract}
{\large
We note that the proof of the no-go theorem of unconditionally secure quantum bit commitment is based on a model which is not universal. For protocols not described by the model, this theorem does not apply. Using unstable particles and a modified double-slit setup, we construct such a protocol and show that it is unconditionally secure. In this protocol, the committer transfers no quantum states to the receiver.}
\end{abstract}



\maketitle


Quantum bit commitment (QBC) is a two-party primitive in quantum cryptography. The security of QBC is of great interest because it can be used as a building block for more complex cryptographic tasks  \cite{Brassard-96,Bennett-91,Kilian88,Crepeau-95}. Unfortunately it is also widely believed \cite{Note} that unconditionally secure QBC is ruled out by a "no-go theorem" put forth in 1997 \cite{Mayers97,LoChau97}. We note that the proof of the theorem is based on a model of QBC. Clearly if the model covers all imaginable QBC protocols, then the theorem can be claimed to be universally valid. We will show that this is not the case.

Conceptually the no-go result of unconditionally secure QBC is puzzling. In physics, we know that if a process is strictly forbidden, it must be ruled out by a law of nature or some symmetry principle which is not broken. For example, electric charge can neither be created nor destroyed because the theory of quantum electrodynamics obeys a global U(1) gauge symmetry.
Another example is the perfect cloning of unknown quantum states. The fact that it is forbidden can be understood by the relativistic requirement that information cannot travel faster than the speed of light.
However as far as we know, unconditionally secure QBC does not violate any known laws or symmetry principles in physics or information theory, so why is it forbidden? In this paper we show that in fact it is not.

For the sake of discussion, a brief review of the subject is in order.
Bit commitment involves two untrusting parties, a committer (Alice) and a receiver (Bob). To begin, Alice commits to a secret bit $b\in\{0,1\}$ which is to be unveiled to Bob at some indefinite later time. To make sure that she cannot change her mind, Alice gives Bob a piece of evidence that can be used to verify her honesty when she unveils. A bit commitment protocol is secure if it satisfies the following two conditions: (1) Concealing: Bob can obtain no information about the value of $b$ before Alice unveils it; (2) Binding: Alice cannot change the committed bit $b$ without being discovered. Furthermore if the protocol remained secure even if Alice and Bob had unlimited computational power, or capabilities limited only by the laws of nature, then it is said to be unconditionally secure.

As is well known, classical protocols are not unconditionally secure. In the quantum version, the evidence that Alice gives to Bob is a quantum state $\rho_B^{(b)}$ which encodes the value of $b$. It is easy to see that if
 \beq
 \rho_B^{(0)}=\rho_B^{(1)}\label{perfect},
 \eeq
then the protocol is concealing. When Alice unveils $b$,
she must provide additional information which,
together with $\rho_B^{(b)}$, will allow Bob to check if she is honest.

What the no-go theorem of unconditionally secure QBC \cite{Mayers97,LoChau97} essentially says is that, using quantum entanglement, Alice can keep all undisclosed classical data undetermined at the quantum level. Then at the end of the commit phase, the quantum state possessed jointly by Alice and Bob can always be regarded as a pure state, $\Psi^{(b)}_{AB}$, such that
 \beq
 \rho_B^{(b)}=
 {\rm Tr}_A~ \ket{\Psi^{(b)}_{AB}}
 \bra{\Psi^{(b)}_{AB}}, \label{reduced}
 \eeq
where the trace is over the Hilbert space controlled by Alice. Then the concealing condition of Eq.(\ref{perfect}), together with a theorem by Hughston $et$ $al.$ \cite{Hughston-93}, imply that $\ket{\Psi^{(0)}_{AB}}$ and $\ket{\Psi^{(1)}_{AB}}$ are related by a unitary transformation $U_A$ on Alice's side:
 \beq
 \ket{\Psi^{(1)}_{AB}} = U_A \ket{\Psi^{(0)}_{AB}}.
 \label{UA}
 \eeq
Since $U_A$ acts on Alice's quantum states only, she can implement it without Bob's help. That means Alice can change her commitment with no risk of being discovered.
Consequently it has been claimed that no QBC protocols can be binding and concealing at the same time.

It should be pointed out that although the above reasoning looks general, it is actually based on a model of QBC. While the model does cover a large class of QBC protocols, it is however not universal. Implicitly this model makes two assumptions: (I) At the end of the commit phase, Bob always possesses a quantum state $\rho_B^{(b)}$ which encodes the bit ($b$) information; (II) The quantum particles involved are stable. Assumption (I) seems reasonable and even unavoidable at first glance, because Bob needs something to bind Alice's commitment. The disadvantage of assumption (I) is that Alice's committing procedure is severely restricted by the concealing condition Eq. (\ref{perfect}). As we saw, this feature is crucial in the proof of the no-go theorem. In order to evade the no-go result, one could imagine a protocol in which Alice transfers no quantum states to Bob at all. On the one hand this feature allows new freedom for Alice's possible actions of commitment, on the other hand it begs the question: If Alice is not required to provide any quantum states to Bob as evidence, what could force her to commit herself during the commit phase? This question can be answered by removing assumption (II) as well. That is, instead of stable particles which last forever, one can employ unstable particles with finite lifetimes, such as neutrons, muons, and {\it etc}. With unstable particles, Alice must commit before the particles disintegrate, or she will have no proper quantum states on which to execute her commit procedure.

In the following, we present a QBC protocol without making assumptions (I) and (II), and prove that it is unconditionally secure. Our protocol is basically a modified double-slit experiment with unstable particles.

%
\begin{itemize}
\item[]\hspace{-0.5cm}{Commit:}
\item[1.]{Bob generates an unstable particle $w$ and sends it towards a double-slit. There are four possible settings at the slits: Both slits are open, the left one is shut, the right one is shut, and both are shut. These settings occur randomly with equal probabilities.}
\item[2.]{For $b=0$, Alice detects the $w$ on a screen at a distance $D$ from the slits, and records its position. For $b=1$, Alice determines which slit (left or right) did the $w$ come through. In either case, Alice must announce whether a $w$ is successfully detected. (Note that if necessary Alice should renormalized her detection probability so that it is independent of $b$.)}
\item[3.]{The above procedure is repeated until $N$ $w$'s have been detected by Alice, where $N$ is some large number. The commit phase terminates at a time $\Delta t\ge 10\tau^w_{1/2}$ after the detection of the last $w$, where $\tau^w_{1/2}$ is $w$'s half-life.}
\item[]\hspace{-0.5cm}{Unveil:}
\item[1.]{Alice unveils $b$ and discloses her detection data. Specifically, for $b=0$ she reveals the position at which each $w$ was detected on the screen, and for $b=1$ she must specify the slit (left or right) through which each of the detected $w$ emerged.}
\item[2.]{Bob checks if Alice's data are consistent with the unveiled $b$. For $b=0$ the double-slit events must combine to form an appropriate interference pattern, while the two sets of single-slit events do not. For $b=1$ the which-slit information for the single-slit events must be accurate, whereas the corresponding double-slit data are random.}
\end{itemize}

We proceed to prove that this protocol is unconditionally secure. First of all, it is trivially concealing, since the only classical information disclosed by Alice during the commit phase is how many $w$'s are detected, and Bob receives no quantum states from Alice.

To prove that our protocol is also binding, we first make two observations: (1) Alice cannot distinguish a double-slit event from a single-slit one with certainty, because the corresponding wave functions are not orthogonal. (2) Alice must finish her commit procedure before the end of the commit phase. If she chose to do nothing, for example, then nearly all of the unstable particles would have disintegrated spontaneously when the commit phase ends. In principle Alice still owns the decay products, but they are useless for cheating purposes. Take neutron ($n$) for example, it decays spontaneously via weak interactions into proton ($p$), electron ($e$), and anti-electron neutrino ($\bar\nu_e$),
 \beq
 n\rightarrow p+e+\bar\nu_e\label{ndecay},
 \eeq
with a half-life of $\tau^n_{1/2}$=608.9 seconds \cite{PDG-22}. After a neutron has decayed, it is no longer meaningful to talk about detecting its position on a screen. In fact, the law of weak interactions dictates that nearly all of the anti-neutrinos would escape detection because their interactions with matter are extremely weak, so weak that they can traverse the whole Earth without being scattered. (At the relevant energy regime, the mean free path of the anti-neutrinos in steel is about ten light years \cite{McFarland08}.) Another viable candidate is muon which has a shorter half-life of $\tau^{\mu}_{1/2}=1.523\times 10^{-6}$ seconds \cite{PDG-22}, its decay mode is
 \beq
 \mu\rightarrow e+\bar\nu_e+\nu_{\mu}\label{mudecay},
 \eeq
where $\nu_{\mu}$ is muon neutrino.

We can now examine classical and quantum cheating strategies:

\noindent (1) Classical cleating: After Alice has determined the which-slit information for $b=1$, the single-slit and double-slit events are not separable, so that a consistent reconstruction of the double-slit interference pattern on the screen is impossible. Similarly, after detecting the positions of the $w$'s on the screen for $b=0$, the resulting pattern is a superposition of single-slit and double-slit events occurring in random order. Again single-slit and double-slit events are not separable, and it is impossible to accurately reproduce the which-slit information for the singlet-slit events. If Alice tries to cheat, she would have to do so by pure guessing, and her success probability obviously vanishes for large $N$. Hence classical cheating is impossible.

\noindent (2) Quantum cheating: Let us assume that Alice can use quantum entanglement to keep private data undetermined until she needs to unveil. However it is obvious that position measurements for $b=0$ and which-slit measurements for $b=1$ produce distinguishable quantum states $\rho_A^{(0)}$ and $\rho_A^{(1)}$ respectively, hence they are $not$ related by a unitary transformation on the ancillas. However the protocol remains concealing because $\rho_A^{(b)}$ belongs to Alice, not Bob. Thus we conclude that quantum cheating is also impossible.

In summary, using a modified double-slit setup and unstable particles, we have constructed a QBC protocol which is not ruled out by the no-go theorem \cite{Mayers97,LoChau97}. This protocol is unconditionally secure because of the following reasons:
(1) Alice does not transfer any quantum particles to Bob, allowing new freedom in her commit procedure.
(2) With unstable particles, Alice must honestly commit (either classically or quantum mechanically) before the end of the commit phase, otherwise she would have no proper quantum states to execute her commitment. (3) The commit procedures for $b=0$ and $b=1$ produce distinguishable quantum states, $\rho_A^{(0)}$ and $\rho_A^{(1)}$ respectively, which are $not$ related by a unitary transformation on the ancillas. (4) Despite the fact that $\rho_A^{(0)}\ne\rho_A^{(1)}$, our protocol remains concealing because Bob has no access to $\rho_A^{(b)}$ which belongs to Alice.


}




\begin{thebibliography}{}
\bibitem{Note} Our discussion is restricted to the non-relativitic regime. See \cite{Kent99} for relativistic effects.
\end{thebibliography}
\end{document}